# An Exploratory Analysis of the Impact of Named Ranges on the Debugging Performance of Novice Users

Ruth McKeever, Kevin McDaid, Brian Bishop
Software Technology Research Centre, Dundalk Institute of Technology,
Dundalk, Ireland
catherineruth.mckeever@dkit.ie, kevin.mcdaid@dkit.ie, brian.bishop@dkit.ie

**ABSTRACT**

*This paper describes an exploratory empirical study of the effect of named ranges on spreadsheet debugging performance. Named ranges are advocated in both academia and industry, yet no experimental evidence has been cited to back up these recommendations. This paper describes an exploratory experiment involving 21 participants that assesses the performance of novices debugging a spreadsheet containing named ranges. The results are compared with the performance of a different set of novices debugging the same spreadsheet without named ranges. The findings suggest that novice users debug on average significantly fewer errors if the spreadsheet contains named ranges. The purpose of the investigative study is to derive a detailed and coherent set of research questions regarding the impact of range names on the debugging performance and behaviour of spreadsheet users. These will be answered through future controlled experiments.*

## 1    INTRODUCTION

Lack of regulation in the financial sector is now more topical than ever. The introduction of legislation such as Sarbanes-Oxley, which enforces financial controls on spreadsheets in businesses in the US, has focused minds on the issue of spreadsheet error. These errors are partially attributed to the distinct lack of professional developers involved in creating spreadsheets, as the majority of spreadsheets are created by the actual users. A further issue is the lack of structured methodologies. One study found that only 6% of development time is spent testing spreadsheets [Baker et al, 2006]. Another [Powell et al, 2007a] found errors in 94% of spreadsheets, and 1-2% of cells. The single largest error found by [Powell et al, 2007b] had an impact of greater than $100 million.

Many attempts have been made to develop tools and best practices that would reduce the high error rates found in spreadsheets, such as WYSIWYT (What You See Is What You Test) [Rothermel et al, 2000], and U-Check [Abraham & Erwig, 2007]. Despite these efforts the instances of material errors continue to occur, both accidentally and intentionally. As an example, in 2005 a sorting error caused aspiring police officers to be incorrectly informed that they had passed an exam [EuSpRIG, 2009].

This research seeks to establish if the use of named ranges could make a spreadsheet easier to understand, and therefore easier to debug. The experiment detailed in this study examines the performance of novice users asked to debug a spreadsheet, seeded with errors, that makes extensive use of named ranges (all the formulas in the spreadsheet use names). The authors also question how frequently named ranges are used in real-world spreadsheets, and examine a repository of sample spreadsheets in order to answer this.

The motivation behind this study is research into how refactoring methods, as used in software engineering, can be applied in practice to support the development of better quality spreadsheets. The goal of the work it to generate tighter research hypotheses and questions to guide future research as to the merits or otherwise of named ranges in spreadsheet technology.

Refactoring is a feature of Agile Software Engineering. It is a technique for cleaning code by making small changes that improve the internal structure of the code, without changing the external behaviour of the program [Fowler M. 1993]. These small changes make errors easier to find, as the code becomes more understandable. Two important Refactoring methods are *Rename Method* and *Replace Magic Number with Symbolic Constant*. *Rename Method* is centred on the principle that a method name should be changed to reflect its purpose. Developers are encouraged to think about what a comment would say about the method, and rename the method accordingly. Likewise, a formula in Excel can be renamed to reflect its purpose. The *Replace Magic Number with Symbolic Constant* method aims to eliminate the unnecessary hard coding of numbers into software, as this practice frequently leads to bugs. Instead the number should be assigned to a variable, which can then be changed in one place instead of throughout the code. Constants in a spreadsheet can be defined in exactly the same way so that they can be changed in one place rather than in every cell that uses them.

Spreadsheet engineering is more closely aligned to agile rather than document driven development processes as utilised in the creation of computer software. It values working software over documentation – often omitting documentation entirely, and can easily respond to change. [Baker et al, 2006] Cite [Cragg and King, 1993] that over 85% of spreadsheets had been modified after their initial implementation and that models are updated an average of 7 times. This is the motivation for using agile practices in an attempt to improve the quality of spreadsheet development.

### 1.2 Overview

This paper has the following layout. Section 2 details the background research beginning with a description of named ranges, an explanation of what the authors regard as quality in spreadsheets, and the research questions that guided this study. Section 3 describes the methodology used to explore the research questions. Section 4 describes the results of the investigative experiment. Section 5 contains a discussion of the results, covering possible causes, the limitations of the study, and the new research questions developed. Section 6 concludes this paper.

## 2  BACKGROUND RESEARCH

### 2.2  Named Ranges

A range is an individual cell, or group of cells. By naming a range it can then be referred to in formulas throughout the spreadsheet in the same way that a variable is named in software code. By giving a range a meaningful name, as one would give to a variable or method in code, it is believed that formulas will become clearer to the user, therefore more understandable and testable. Without meaningful range names the user must remember the meaning of a cell named, for example "H79" and then check for its occurrence in formulas throughout the workbook. E.g. the formula "netIncome =

grossIncome – tax", at first glance is far more understandable than "E80 = A40 – D69". To name a range the developer simply highlights the range and enters a suitable name in the name box (the box above and to the left of the worksheet that normally contains the reference of the cell that is currently in focus).

The inclusion of the Name Manager in the Excel 2007 ribbon greatly improves the ease with which the user can add, modify and delete names. It also provides the facility for the user to sort and filter the names, and provides a quick insight to what each range refers. While typing a name in the name box is a simple way of creating a range name, it is not possible to change either the name, or the values to which it refers, in this location. The Name Manager tab gives the developer quick and easy access to these type of modifications. It also allows the developer to associate further information with the name, in the form of a 255-character comment.

To go to a named range in a workbook the user can either click on the name box, or press *F5* on the keyboard, and select which range they wish to go to. This brings the focus directly to the named range. The user can also insert a list of all the named ranges into a worksheet by pressing *F3* and choosing *Paste List*. A developer can name a single cell, a group of cells, a constant, or a formula. Used properly names can be a powerful tool with many properties and uses, including the following:

*Inserting a reference:* By naming a cell or group of cells you can insert that range elsewhere in the spreadsheet simply by referring to the name. Because names are set as absolute by default, the reference will not change. If the original cells are subsequently moved the name will still refer to the same values, hence the inserted values will remain correct. If the values to which the reference refers change, then the inserted values will also be updated.

*Different levels:* Names are most commonly used at workbook level, but they can also be declared at sheet level. This can be useful so that one name can be used refer to the same range of cells on several sheets.

*Absolute and relative referencing:* Names are absolute by default, but when creating a named formula the developer has the option of using absolute or relative referencing. If the developer wishes to use relative cell referencing it is important that they are in the cell that the formula will initially be used in, as it is from this cell that the reference will refer.

*Constants:* Constants can be named without needing a cell reference. In the Name Manager dialog box the developer simply defines a name, and in the *Refers To* field enters a value rather than a range or cell reference.

*Validation:* Named ranges can be used for data validation by naming a range of allowed values and then using this name as the source for the validation list.

*Dynamic named ranges:* Dynamic named ranges allow the developer to name a range where the size of the range is not known from the outset, or if the size may change. This is implemented with use of the OFFSET and COUNT functions to calculate how many values are contained in a range and then setting the name accordingly.

Other technologies have been developed to support the user in the management of range names. One example is OPERIS Analysis Kit (OAK) [OPERIS, 2009] which includes features to modify names to correct misspellings, apply and delete multiple range names and, most interesting in the context of the results we will present later, the ability to replace range names with cell references.

### 2.3 Quality in Spreadsheets

Quality of spreadsheets in this study refers to reliability, understandability, testability and extendibility. The reliability of a spreadsheet is essentially the accuracy of the data that it produces, and is compromised by the errors found in approximately 94% of spreadsheets. Understandability refers to how easily a user or auditor can make sense of the spreadsheet, and is fundamental to the implementation of Sarbanes-Oxley. Testing is crucial if the reliability of the spreadsheet is to be proven, yet is next to impossible if the spreadsheet is not understandable. Extendibility relies on the previous characteristics, yet is vitally important considering how frequently spreadsheets are reused and remodelled.

### 2.4 Research Questions

The review of literature related to the use of Named Ranges in spreadsheets was guided by the following questions:

- Research Question 1 (RQ1): Do industrial, academic, standards and training organisations advocate the use of named ranges, and if so, how should they be used and why?
- Research Question 2 (RQ2): To what extent and in what way are range names used in practice?

**RQ1: Do industrial, academic, standards and training organisations advocate the use of named ranges, and if so, how should they be used and why?**
The SSRB (Spreadsheets Standards Review Board) [Hutchens, 2005], Microsoft [Microsoft Corporation, 2006], and Read & Batson (1999) advise the use of named ranges. In their paper for IBM [Read & Batson, 1999] the authors state that "allocating meaningful range names to areas or cells within a spreadsheet can speed up the development process, make the model easier to understand and reduce the risk of errors made by referring to the wrong cell." They also suggest naming constants rather than hard-coding them into cells, as this makes them easier to change, and recommend referring nearby cells by cell reference and far-away formulas by range name. In their 2006 white paper [Microsoft Corporation, 2006], Microsoft recommend "named ranges to reduce errors and increase formula readability." The SSRB, in their Best Practice Spreadsheet Modelling Standards [Hutchens, 2005], describe detailed naming principles, including four naming conventions related to range naming, "Every range name in the workbook should describe the content or use of the range being named". They provide a list of prefixes that should be used with different types of names.

Range naming is recommended by many websites devoted to spreadsheet advice [Pearson, 2007], [Mr Excel, 2007], [Ozgrid, 2008] and [Johnson, 2008], and by Microsoft in their online Excel documentation [Microsoft, 2008]. As far back as 1985 the Journal of Accountancy ran an article [Bromley, 1985] in which the author states that defining names for cell ranges "reduces the probability of cell reference errors caused by moving a column or row to another location." [Bewig, 2005] advocates the proper construction of range names for eliminating the problem of referring to the wrong cell while constructing formulas, and states that "well-chosen names are the first and best form of documentation."

In an article on Spreadsheet Accuracy Theory [Kruck & Sheetz, 2001], the authors propose that naming ranges can help the developer comprehend cell meaning "As cells become more highly interconnected, the developer spends inordinate time trying to remember the meaning of the cell, and she/he is distracted from using it effectively. Naming of ranges [Miller 1989] and structure [Ronen Palley Lucas 1989] helps with this problem." The author sums this research by stating "simple policies requiring separation of data areas and user interface areas, requiring cell naming, limits on formula length, or documentation of tests conducted would result in more accurate spreadsheets".

One survey was conducted of spreadsheet users in Australia [Hall M. 1996], which found that 60% of the surveyed participants used named ranges. The author then stated that 75% of the participants said that they should have used named ranges, however there was no elaboration as to why they felt that they should be used.

Range naming is not without its detractors. The authors of [Panko & Ordway, 2005] warn against the dangers of a range name referring to an incorrect range, and therefore appearing correct when it is not. They state "although the research findings are not clear on this issue, using range names should be considered potentially dangerous until research on using range names is done."

The author of [Blood, 2002] advises against using named ranges, stating that they are unnecessary if the developer has designed the layout well. Three main disadvantages of range names are set out in this paper. Firstly, very descriptive range names lengthen simple formulas, which should already have their function made clear by a row label. Secondly, range names hide the location of the cells to which they refer, and thirdly they create "ghost" links when sheets are copied to a different workbook.

Despite this conflicting advice from spreadsheet experts, there is no academic research at present that examines whether the use of names in spreadsheets increases quality.

**RQ2: To what extent and in what way are named ranges used in practice?**
To establish how named ranges are currently used, a quantitative evaluation of existing spreadsheets was developed. This analysis was carried out on the EUSES Corpus [Fisher M. 2005], a repository of 5606 real-world sample spreadsheets, including 4498 unique spreadsheets, available to spreadsheet researchers. The authors found that 51% of these spreadsheets contained names, however not all these names referred to ranges. When studied in more depth it was found that of those spreadsheets that do contain names, 46% use these names exclusively for defining print areas, and only 2% of them use names in formulas. The remainder of the spreadsheets contained names that referred to tables, names that linked other workbooks or databases, appeared to either have no function or had their reference deleted.

### 2.5 Investigative Research

After reviewing the literature, it was decided to examine the impact of range names on debugging performance. An investigative experiment was designed to explore how novice users perform debugging a spreadsheet seeded with errors that was developed using range names wherever possible. This experiment is described in the following section.

### 3 METHODOLOGY

The experiment used to evaluate the hypothesis was based on research previously conducted by the authors of [Bishop & McDaid, 2007]. It involves asking a group of novice spreadsheet users to examine a spreadsheet and to correct any errors that they found. The experiment described in [Bishop & McDaid, 2007] is used as the comparative group for this study. This approach has several limitations, which are discussed in section 5.

**3.2 Original Experiment**

This original experiment was described in [Bishop & McDaid, 2007] and compared the debugging capabilities of expert versus novice spreadsheet users. It was based on another study conducted by [Howe & Simpkin, 2006] and consisted of a spreadsheet containing three sheets: *Payroll*, *Office Expenses* and *Projections*. *Payroll* calculates the payroll expenses for a typical week, *Office Expenses* sums office expenses for the first quarter and estimates the same for the remainder of the year, and *Projections* estimates the total expenses, both office and payroll, for the next 5 years. This spreadsheet was seeded with 42 errors, and contained no range names.

While the group debugged the spreadsheet, T-CAT, a "time-stamped cell activity tracking tool", [Bishop B., McDaid K. 2008] ran in the background recording data about each cell click. This tool was developed as a macro in VBA in order to record the interaction between the participant and the spreadsheet with minimal intrusion. This data recorded included each cell entered, at what time each cell was entered, and what changes were made. The resulting data is printed to a hidden worksheet when the spreadsheet is closed and can then be analysed to examine each participants' interaction with the spreadsheet.

The errors seeded in this trial can be divided into four categories: *Clerical/Non material*, *Rule Violation*, *Data Entry* and *Formula*. There were four *Clerical* errors such as spelling mistakes, and four *Rule Violation* errors where the data in the spreadsheet violated the company policy as detailed in the instructions sheet. There were 8 *Data Entry* errors, for example where a number was entered incorrectly. The *Formula* errors were further divided into four sub-categories: *logic*, *cell*, *range* and *remote*. *Logic* errors consisted of illogical formulas, and instances where a formula was expected but a number used instead. *Cell* reference errors consisted of errors where an incorrect individual cell was referenced while *range* errors consisted of errors where an incorrect group of cells were referenced. *Remote* reference errors occurred where an incorrect reference to a different sheet was used.

**3.3 Experiment on Named Ranges**

For our study the original structure and format of the spreadsheet was retained, as were the 42 seeded errors. The only change was the implementation of named ranges. Individual cells were named rather than arrays, because of the difficulties faced when attempting to change a value in an array.

This experiment group consisted of 21 students in their second year of a computing degree in Dundalk Institute of Technology, who had a year earlier taken a class in spreadsheet basics such as formulas and charts. This experiment was carried out in accordance with the ethics policy in Dundalk Institute of Technology. The participants were first given a tutorial on naming ranges, composed of a presentation and a practical exercise that asked them to create, edit, use and delete range names. The authors monitored their performance in this introductory task and were satisfied that they were able to use named ranges to the extent required to perform the experiment. As shall be shown, their performance in the task clearly supports this assumption. When this was completed satisfactorily, they were asked to open up the experimental spreadsheet and begin the trial.

The group consisted of a mixture of Irish and international students. There was some concern that the meaning of the range names might not be clear to the students for whom English is not their first language. The initial analysis separated the students but as there was no significant difference in performance between the different nationalities, the following results are based on the group's performance as a whole.

Each student was given an instructions sheet detailing the rules and assumptions that the data in the spreadsheet was expected to follow. They were asked to correct the errors directly on the spreadsheet. The students were not given a time limit for the trial, although it was expected that they would take no longer than an hour.

## 4   RESULTS

The results showed that the students corrected on average 47% of the errors seeded in the spreadsheet. This is approximately 11% less than the comparative group. Unsurprisingly, the only statistically significant difference in performance relates to the debugging of formula errors, where the experiment group corrected only 44% of errors, while the control group corrected 63%.

| Error Type | No of Seeded Errors | % Corrected by Experiment Group | % Corrected by Control Group | Experiment Compared to Control |
|---|---|---|---|---|
| Clerical/Non-Material | 4 | 11% | 11% | 0% |
| Rule Violation | 4 | 63% | 65% | -2% |
| Data Entry | 8 | 64% | 63% | 1% |
| Formula | 26 | 44% | 63% | -19% |
| **Total** | **42** | **47%** | **58%** | **-11%** |

**Figure 1 - Error Correction Results**

The similar results in the *Clerical*, *Rule Violation* and *Data Entry* categories indicate that the control group and experiment group are comparable in terms of ability and knowledge. As the experiment group performed significantly worse in the formula errors, these were examined further to establish which type of formula errors posed the most problems.

| Error Sub-Type | No of Seeded Errors | % Corrected by Experiment Group | % Corrected by Control Group | Experiment Compared to Control |
|---|---|---|---|---|
| Logic | 9 | 54% | 63% | -9% |
| Cell | 7 | <u>39%</u> | <u>68%</u> | <u>-29%</u> |
| Range | 7 | <u>47%</u> | <u>71%</u> | <u>-24%</u> |
| Remote | 3 | 19% | 28% | -9% |
| **Total** | **26** | **44%** | **63%** | **-19%** |

**Figure 2 - Formula Error Correction Results**

While the experiment group performed worse in all the formula sub-categories, the greatest difference was with *Range* and *Cell* reference errors, and these were the only groups where the results were statistically significant at a 5% level based on a non-parametric rank sum test. In each case of these types of error the control group performed better than the experiment group.

These are the precise category of error that range names were expected to reduce, as these errors are all due to either an incorrect name used in a formula (e.g. the formula in *Payroll F11* is "=GriffinPayRate * HartfordRegularHours" instead of "=GriffinPayRate * GriffinRegularHours"), or a name referring to the wrong cell or range (e.g. the formula in *Office Expenses F18* is "=SUM(FixedYearEst)", but this name refers to an incorrect range of cells).

## 5 DISCUSSION

### 5.2 Causes

The results were a surprise to the authors given the widespread opinion that range names improve the understandability and reliability of spreadsheets. We next investigate three possible explanations.

**High cognitive load**
Because the participants were debugging a spreadsheet that they had not personally developed, it was thought that they would have trouble remembering what each name referred to, and might spend additional time switching back to check this, especially if the range used was located on a different sheet. They would have to complete two checks, one to see if the correct range name was used, and another to see if the name referred to the correct range. This would result in a higher cognitive load and thus may explain the poor performance. However, the students were taught how to click on a name in the name box (or by pressing the F5 key) and it would bring them directly to that range. Unfortunately it is unclear how many of them used that facility, as this was not recorded by T-CAT.

**Too much confidence in names**

Because the students were given a tutorial on naming ranges before the experiment commenced, it is possible that they saw the benefits of names and therefore expected them to be correct. This is difficult to prove, but it appeared that on a students' initial reading of a formula, if the first range name appeared to be correct then they did not inspect any further. For example, an error contained in *Payroll F9* occurred when the working hours of one employee were multiplied by the pay rate of a different employee. Only 14% of the experiment group corrected this error, in comparison with 65% of the control group. The erroneous entry was "=EnglebertPayRate*DanielsRegularHours". The correct entry should have been "=EnglebertPayRate*EnglebertRegularHours". 19% of the experiment group, and 65% of the control group corrected a similar error on the same sheet. It seems that when the participants saw that the first name appeared correct (to the extent that the name was what they expected to see) they presumed that the formula was correct.

The experiment group performed significantly (23%) better in one *Formula Logic* error, where a number had been used instead of a formula. This indicates that the error was immediately obvious because it contained no names, and was not consistent with other cells that performed the same task.

**Did not understand the error or did not know how to correct it**
One threat to the validity of this experiment was the possibility that the participant might not fully understand the concept of names. As they were second year computing students they understood the concept of variables. Before the participants began the trial each one watched a presentation and completed a task that taught them how to use range names. This task involved creating new named ranges, creating names to refer to constants, using these names in formulas and editing existing formulas so that they used names instead of cell references. Only when the authors were satisfied that the students could carry out these tasks were they asked to complete the experiment. This ensured that they understood the concept of names.

To investigate whether there was evidence that participants had insufficient knowledge to utilise range names, we examined in detail how each participant handled each range formula error, for which the debugging performance was comparatively low. In some cases it was thought that the subject might have been able to identify the error but unable to repair it. We considered that this would result in numerous unsuccessful attempts at debugging the cells. As Figure 3 shows, while up to 7 attempts were made to correct range errors, of those who attempted these errors the majority succeeded in correcting them. In two cases a participant failed after one attempt, and in one case a participant failed after two attempts. This reinforces the authors' belief that the participants had sufficient knowledge about naming in order to correct the errors.

| | **Range Errors** | | | | | | | | | | |
|---|---|---|---|---|---|---|---|---|---|---|---|
| **Error Location** | **Corrected on Attempt No.** | | | | | | | | **Failed on Attempt No.** | | |
| | 1 | 2 | 3 | 4 | 5 | 6 | 7 | Total | 1 | 2 | Total |
| Payroll G16 | 8 | 1 | | | | | | 9 | | | 0 |
| Payroll H16 | 6 | 1 | 1 | | | | 1 | 9 | | | 0 |
| Payroll I10 | 5 | 5 | 2 | 1 | | | | 13 | | | 0 |
| Payroll I14 | 8 | 2 | 2 | | 1 | | | 13 | | | 0 |
| Office Expenses F10 | 6 | | 1 | | | | | 7 | 1 | | 1 |
| Office Expenses F18 | 5 | 1 | | 1 | 1 | 1 | | 9 | 1 | | 1 |
| Projections G22 | 4 | 1 | 2 | 2 | | | | 9 | | 1 | 1 |

**Figure 3 Range Reference Errors**

Figure 3 shows the location of the range errors, how many participants corrected, or failed to correct, each error, and after how many attempts.

An error in *Office Expenses F18* referred to an incorrectly defined name on a different sheet. 43% of the experiment group, and 74% of the control group corrected this. This cannot be attributed to a lack of knowledge regarding how to fix the error, as of the 10 participants who attempted to fix the error, 9 succeeded. This was the only error where a range was incorrectly defined.

### 5.3 Limitations

An alternative approach might have been to randomly split the 21 participants into experimental and control groups. However, the small number of participants made this unworkable in terms of deriving statistically significant results. Instead, we argue below that the performance of the entire group can be compared with the group from the previous study.

Figure 4 compares the results of three trials: the experiment described in this paper (Group A), the original experiment used as the control group (Group B), and the control group from another experiment (Group C). The novice users in the original trial consisted of thirty-four second year accounting and finance students. The third trial had been carried out on fourth year Software Development students as part of another study [Bishop B., McDaid K. 2008].

 

| Error Type | No of Seeded Errors | % Corrected by Group A | % Corrected by Group B | % Corrected by Group C |
|---|---|---|---|---|
| Clerical/Non-Material | 4 | 11% | 11% | 13% |
| Rule Violation | 4 | 63% | 65% | 66% |
| Data Entry | 8 | 64% | 63% | 66% |
| Formula | 26 | 44% | 63% | 63% |

**Figure 4 Comparison of Results by Group**

Because of the consistencies between the groups of students that had already partaken in this experiment, and as this study is exploratory, the authors deemed the novice users from the original experiment to be suitable for use as a control group.

Another major limitation of this study is that its sole focus is on novice users. The effect of named ranges on expert users cannot be presumed, as it has been proven that professionals are significantly better than novices at correcting particular types of errors [Bishop & McDaid, 2007].

The students chosen for this study were computing students and therefore already understand the concept of variables. Spreadsheet users from a different background are unlikely to understand this principle so easily.

### 5.4 Future Work

The results suggest clearly that range names can have a detrimental affect on spreadsheet debugging. It is important to use the results to provide more focused and coherent research questions that can be answered through further PhD study by the first author.

To that end, and based on the findings of the experiment, new hypotheses are being considered such as:

*A spreadsheet that contains range names in formulas will be more difficult to inspect and correct than a spreadsheet that does not use names in formulas.*

Taking into account the limitations noted above, new research questions are under consideration relating to the hypothesis:

- Does the use of named ranges give the user (false) confidence in the accuracy of the spreadsheet?
- Does the use of named ranges increase the cognitive load on users debugging a spreadsheet?
- Do professional spreadsheet users face the same difficulties as novices when debugging a spreadsheet that contains named ranges?

In order to answer these new research questions, a new experiment is planned. This experiment will be based on the original experiment, but will be more carefully developed to ensure a balance between errors in named ranges, and errors in unnamed ranges. The range errors will also be balanced between those where a wrong name is used, and those where the name refers to the wrong range. In order to answer RQ3 and RQ4, a more detailed analysis of the time each participant spends in each cell relating to the errors will be carried out. In order to answer RQ5, the experiment will be performed on both novice and professional spreadsheet users.

Based on the limitations found in the exploratory study it has been decided that computing students will not be used for this study, as their domain knowledge of variable names may have an affect on their use of range names. The group used in this study will be divided randomly into a control group and an experiment group. If possible, the T-CAT tool will be modified to record if the user makes use of the 'Go To Name' functions of the workbook.

A qualitative survey should be carried out on expert users to investigate their attitude to, and experience of, range names. An examination on the effect of naming ranges while developing a spreadsheet should also be undertaken, however no plans have been made regarding this yet.

## 6   CONCLUSION

While conventional wisdom might suggest that named ranges have a positive effect on spreadsheets, this work suggests otherwise, as the results indicate that named ranges may lead to a reduction in debugging performance of novice spreadsheet users.

These results do not, however, prove that named ranges will always have a negative effect on spreadsheets. Indeed it may have been the overuse of names, or the lengthening of formulas, that resulted in the poor performance by the experiment group, as the quantity of names used may well have distracted from their intrinsic worth. These results indicate that extreme care must be taken when advising users about developing spreadsheets. Development practices that might seem advantageous in principle can have unexpected consequences.

In essence, this paper supports, through a quantitative study, the possibility that range names may have a detrimental impact on spreadsheet debugging. Furthermore, it stresses the need for a larger and more tightly controlled and focused empirical study to answer this important question.

**REFERENCES**

Abraham, R. & Erwig, M. (2007) "UCheck: A Spreadsheet Type Checker for End Users" Journal of Visual Languages and Computing, Vol.18 (1) Feb. 2007, 71-95.

Baker, K., Fostor-Johnson, L., Lawson, B. & Powell, S. (2006) "A Survey of MBA Spreadsheet Users", [Online] Available: http://mba.tuck.dartmouth.edu/spreadsheet/product_pubs.html [March 2009]

Bewig, P. L. (2005) "How Do You Know Your Spreadsheet Is Right? Principles, Techniques and Practices of Spreadsheet Style", [Online] Available: http://www.eusprig.org [March 2009]

Bishop, B. & McDaid, K. (2007) "An Empirical Study of End-User Behaviour in Spreadsheet Error Detection & Correction", http://arxiv.org/abs/0802.3479 Proceedings of the European Spreadsheet Risk Interest Group Conference, 2007.

Bishop, B. & McDaid, K. (2008) "Spreadsheeet End-User Behaviour Analysis", http://arxiv.org/abs/0809.3587 Proceedings of the European Spreadsheet Risk Interest Group Conference, 2008.

Blood, A. T., (2002) "Elements of Good Spreadsheet Design: Spreadsheet Design Notes" [Online] Available: http://www.xl-logic.com/modules.php?name=Content&pa=showpage&pid=2 [May 2009]

Bromley, R. G. (1985) "Template Design and Review: how to prevent spreadsheet disasters" Micros in Accountancy, Journal of Accounting, December 1985, 134-145

EuSpRIG, "Spreadsheet mistakes – news stories", [Online] Available: http://www.eusprig.org/stories.htm [Mar. 24, 2009]

Fowler, M. "Refactoring: Improving the Design of Existing Code". New York: Addison-Wesley, 1993.

Fisher, M., & Rothermel, G. (2005 ) "The Euses Spreadsheet Corpus: A Shared Resource for Supporting Experimentation with Spreadsheet Dependability Mechanisms." WEUSE05: 1st Workshop on End-User Software Engineering, 2005, pp.47-51.

Hall, M. J. J (1996) "A Risk and Control-Oriented Study of the Practices of Spreadsheet Application Developers," Proc. of the 29th Annual Hawaii International Conference on System Sciences

Howe, H. & Simpkin, M.G. (2006) "Factors Affecting the Ability to Detect Spreadsheet Errors", Decision Sciences Journal of Innovative Education, January 2006, Vol.4(1)

Hutchens, M. (2005) "Best Practice Spreadsheet Modelling Standards" [Online] Available: http://www.ssrb.org/best_practice_spreadsheet_modelling_standards_download.html, [Oct. 22, 2008].

Johnson, L. (2008) "Naming Ranges, Constants and Cells in Excel: The Whys and Hows." [Online] Available: http://personal-computer-tutor.com/names.htm, [Dec. 8, 2008]

Kruck, S. E., & Sheetz, S. D. (2001) "Spreadsheet Accuracy Theory" Journal of Information Systems Vol.12(2) 2001, 93-108

Microsoft Corporation (2006, April) "Spreadsheet Compliance in the 2007 Microsoft Office System" White Paper

Microsoft (2008) "Define and Use Names in Formulas" [Online] Available: http://office.microsoft.com/en-gb/excel/HA101471201033.aspx?pid=CH100648431033, [Nov. 18, 2008].

Miller, S. E. (1989) "8 ways to avoid worksheet errors." Lotus, pp. 50-53.

Mr Excel (2007, November) "Episode 626 – VBA Naming Ranges." [Online] Available: http://www.mrexcel.com/Excel_VBA_Naming_Ranges_training.html [Dec. 8, 2008]

OPERIS Analysis Kit [Online] Available: http://www.operisanalysiskit.com/ [June 2009]

Ozgrid (2008) "Lesson 23 – Using Named Ranges in Excel as an Alternative to Cell References." [Online] Available: http://www.ozgrid.com/Excel/free-training/excel-lesson-23-basic.htm, [Nov. 25, 2008]

Panko, R. R. & Ordway, N. (2005) "Sarbanes-Oxley: What About all the Spreadsheets? Controlling for Errors and Fraud in Financial Reporting" http://arxiv.org/abs/0804.0797 Proceedings of the European Spreadsheet Risk Interest Group Conference, 2005.

Powell, S., Baker, K. & Lawson, B. (2007a) "Errors in Operational Spreadsheets." [Online] Available: http://mba.tuck.dartmouth.edu/spreadsheet/product_pubs.html [March 2008]

Powell, S., Baker, K. & Lawson, B. (2007b) "Impact of Errors in Operational Spreadsheets." [Online] Available: http://mba.tuck.dartmouth.edu/spreadsheet/product_pubs.html [March 2009]

Pearson, C. H. (2007) "Working with Named Ranges in Excel", [Online] Available: http://www.cpearson.com/Excel/named.htm, Sep 6, 2007 [Nov. 25, 2008].

Read, N. & Batson, J. (1999, April) "Spreadsheet Modelling Best Practice." [Online] Available: http://www.eusprig.org/smbp.pdf, [Oct. 22, 2008].

Ronen, B., Palley, M. A., & Lucas, H. C., Jr. (1989) "Spreadsheet analysis and design." Communications of the ACM, pp. 84-93.

Rothermel, K. J., Cook, C. R., Burnett, M. M., Schonfeld, J. T., Green, R. G. & Rothermel, G., (2000) "WYSIWYT Testing in the Spreadsheet Paradigm: An Empirical Evaluation" Proceedings of the 22nd International Conference on Software Engineering, June 2000, 230-239.